\documentclass{elsart3p}

 \usepackage{graphicx}

\usepackage{amssymb}

\begin{document}

\begin{frontmatter}

\title{Effects of the Hybridization on the Fermi Surface of an Extended $d-p$ Hubbard Model}

\author[aff1,aff2]{E. J. Calegari\corauthref{cor1}}
\ead{eleonir@mail.ufsm.br}
\corauth[cor1]{}
\author[aff1]{S. G. Magalhaes}
\author[aff3]{A. A. Gomes}
\address[aff1]{Laborat\'orio de Mec\^anica Estat\'{\i}stica e Teoria da Mat\'eria Condensada, Departamento de 
F\'{\i}sica, UFSM, 97105-900, Santa Maria, RS, Brazil}
\address[aff2]{Instituto de F\'{\i}sica e Matem\'atica - UFPEL, Caixa Postal 354, 96010-900, Pelotas, RS, Brazil}
\address[aff3]{Centro Brasileiro de Pesquisas F\'{\i}sicas, Rua Xavier Sigaud 150,
22290-180, Rio de Janeiro, RJ, Brazil}


\begin{abstract}
The Fermi surface (FS) of an extended $d-p$ Hubbard model is investigated by   
a two-pole approximation in both situations with hole and electron doping.
Using the factorization procedure proposed by Beenen and Edwards,
superconductivity with singlet $d_{x^2-y^2}$-wave 
pairing is considered. The effects of the $d-p$ hybridization on the FS
are the main focus in the present work. Nevertheless, the asymmetries between the 
hole- and electron-doped regimes and the effects of
doping and Coulomb interaction on FS are also investigated. 
Particularly, it is shown that the crossover from hole-like to 
electron-like FS is deeply affected by the $d-p$ hybridization in the hole-doped case.
It has been verified that the effect of the hybridization is very pronounced 
around the saddle points $(0,\pm\pi)$
and $(\pm\pi,0 )$, where the intensity of the superconducting order parameter is maximum 
in the particular case of $d_{x^2-y^2}$-wave symmetry. 
In the electron-doped case, the crossover in the FS is not verified. 
The doping dependence of the FS topology in the
hole- and electron-doped regimes is in agreement with recent experimental ARPES results 
for La$_{2-x}$Sr$_{x}$CuO$_4$ (hole doping) and Nd$_{2-x}$Ce$_x$CuO$_4$ (electron doping).
\end{abstract}

\vspace{0.8cm}
\end{frontmatter}


Almost two decades after the discovery of the high-temperature superconductors (high-$T_c$), 
the theoretical description of this phenomenon still represents a challenge for 
the physicists. The study of the Fermi surface and the dispersions of bands is essential
to better understand the superconducting and the normal properties of the high-$T_c$'s \cite{borisov}. Furthermore, the investigation of the asymmetries between hole- and electron-doped regimes may contribute to clarify the mechanisms of superconductivity in these materials \cite{Guo}. Due to the strong correlations at the Cu-sites,
the one-band Hubbard model has been largely used to describe such systems. Nevertheless, the fact that
the oxygen sites may be occupied by holes when the system is doped suggests that a model which also takes into account the oxygen can be more adequate to treat these systems in the doped regime \cite{calegari2005EB}.
In the present work, the two-pole approximation \cite{Roth} has been used to study the FS associated with
an extended $d-p$ Hubbard model. Superconductivity with $d_{x^2-y^2}$-wave symmetry is considered by using the 
factorization procedure proposed by Beenen and Edwards \cite{beenen}.
The Hamiltonian model is an improved version of the model studied in reference \cite{calegari2005EB}. It is given by:
\begin{eqnarray}
H&=&\sum_{\langle i\rangle j,\sigma }\left\lbrace \left[ ( \varepsilon_{d}-\mu)d_{i\sigma}^{\dag}d_{j\sigma }
+(\varepsilon_{p}-\mu)p_{i\sigma }^{\dag}p_{j\sigma}\right]\delta_{ij}\right. \nonumber\\ 
& &\left. ~~~+t_{ij}^{d}d_{i\sigma}^{\dag}d_{j\sigma }+t_{ij}^{p}p_{i\sigma }^{\dag}p_{j\sigma }+t_{ij}^{pd}( d_{i\sigma}^{\dag}p_{j\sigma +}p_{i\sigma }^{\dag}d_{j\sigma })\right\rbrace \nonumber\\ \nonumber\\
& &~~~+U\sum_{i}n_{i\uparrow}^{d}n_{i\downarrow}^{d}
+\sum_{\langle\langle i\rangle\rangle j,\sigma }(t_{ij}^{ld}d_{i\sigma
}^{\dag}d_{j\sigma }+t_{ij}^{lp}p_{i\sigma }^{\dag}p_{j\sigma }) \nonumber\\
\label{eq1}
\end{eqnarray}
where $\mu$ is the chemical potential. The symbols $\langle ...\rangle$ $\left(\langle\langle ...\rangle\rangle\right)$ denote the sum over the first(second)-nearest-neighbors of $i$. The hopping to the second-nearest-neighbors is necessary to describe correctly the FS topology, mainly in the electron-doped regime. The quantity $U$ stands for the local Coulomb interaction between two $d$-electrons with opposite spins. 
The Green's functions necessary to treat the problem are obtained following the standard Roth's procedure \cite{Roth}. 
In order to include superconductivity and $d-p$ hybridization, in the present work,
the resulting Green's functions consist of a five-pole approximation \cite{calegari2005EB}:
\begin{equation}
G_{\bf{k}\sigma }(\omega)=\sum_{s=1}^5\frac{Z_{\bf{k}\sigma }^{(s)}}
{\omega-E_{s\bf{k}\sigma}}
\label{G11}
\end{equation}
with each pole corresponding to a quasi-particle band $E_{s\bf{k}\sigma}$. The quantity $Z_{\bf{k}\sigma }^{(s)}$
stands for the spectral weight \cite{calegari2005EB}. The quantities $E_{s\bf{k}\sigma}$ and $Z_{\bf{k}\sigma }^{(s)}$
are obtained as in reference \cite{calegari2005EB}.

\begin{figure}[t!]
\begin{center}
\leavevmode
\includegraphics[angle=-90,width=8cm]{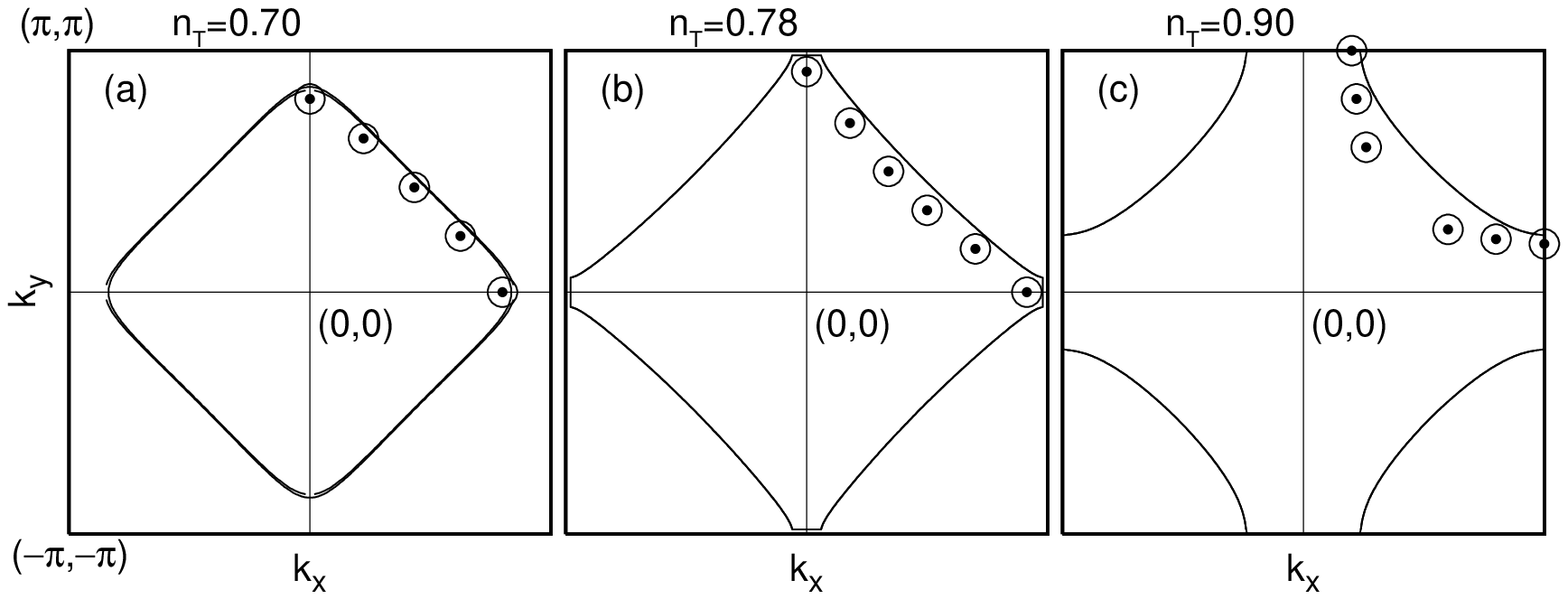}
\end{center}
\begin{center}
\leavevmode
\includegraphics[angle=-90,width=3.5cm]{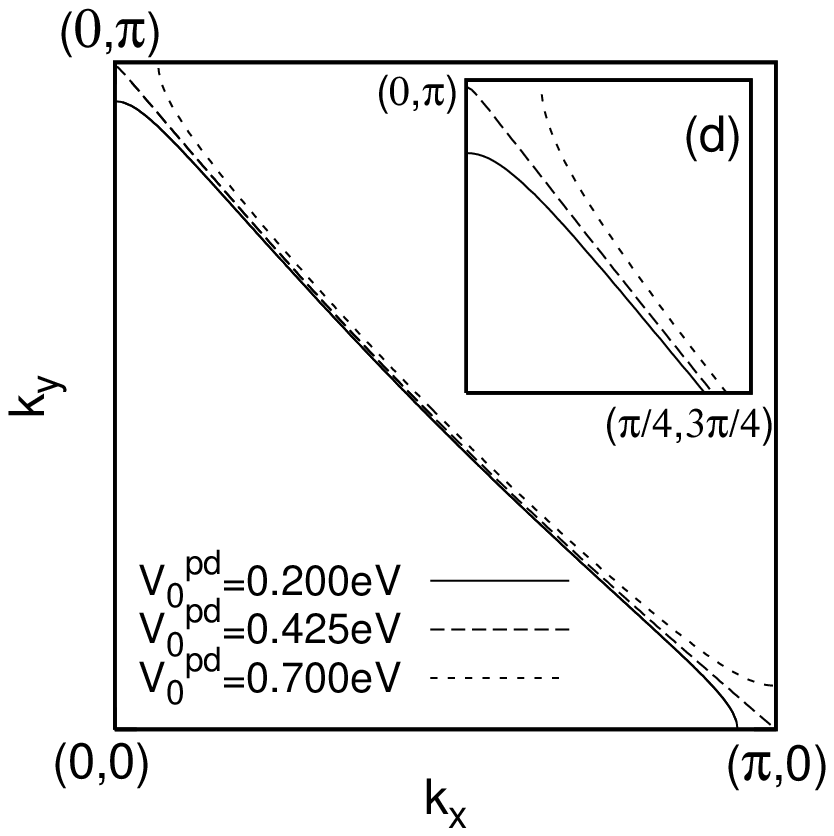}
\leavevmode
\includegraphics[angle=-90,width=3.5cm]{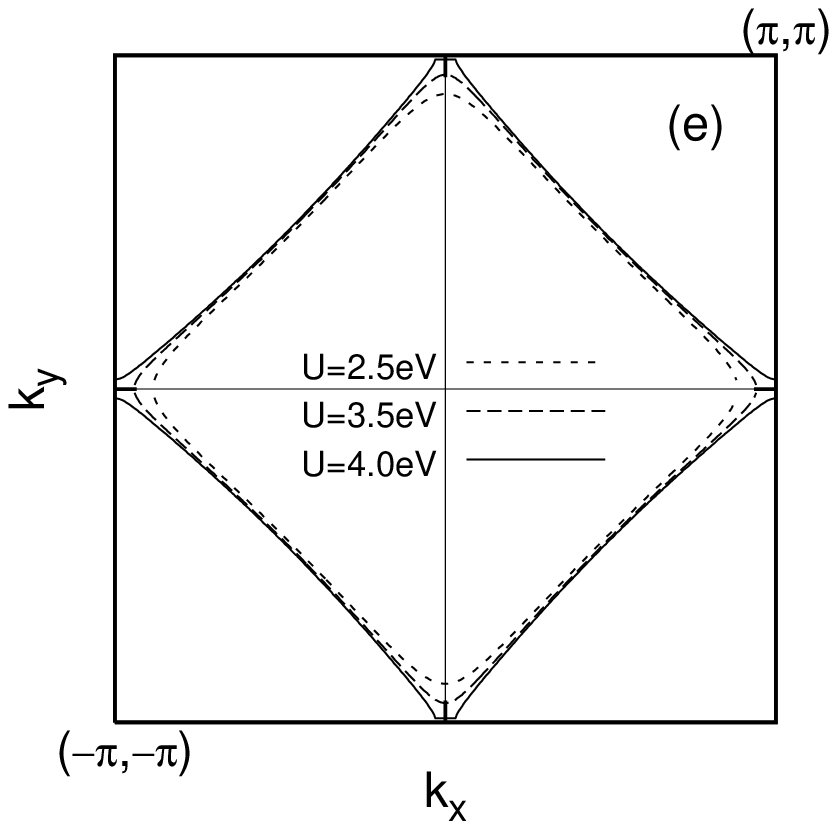}
\end{center}
\caption{(a), (b) and (c) show the evolution of the Fermi surface for different occupations $n_T$ in the hole doping 
     regime with $k_BT=0.0011$eV. The model parameters are $U=3.5$eV, $V_0^{pd}=0.2$eV, $\varepsilon_{p}-\varepsilon_{d}=3.6$eV, $t^d=-0.5$eV, $t^p=-0.7$eV,
     $t^{ld}=0.04$eV and $t^{lp}=0$. The symbols $\odot$ show the experimental data for La$_{x-2}$Sr$_x$CuO$_4$ 
     taken from Ref. \cite{Ino2}.
(d) Fermi surface for $n_T=0.76$ and different values of hybridization. 
(e) Fermi surface for $n_T=0.76$ and different values of Coulomb interaction $U$.}
\label{figure:FSh}
\end{figure}

The figures \ref{figure:FSh}(a), \ref{figure:FSh}(b) and \ref{figure:FSh}(c) show Fermi surfaces for different 
hole-doping. The symbols $\odot$ show
the experimental ARPES results for La$_{2-x}$Sr$_x$CuO$_4$ (LSCO) \cite{Ino2}.
The figure \ref{figure:FSh}(a) shows that, in the hole overdoped regime $(n_T=0.70$, with $n_T=n_{\sigma}^d+n_{-\sigma}^d )$,
the Fermi surface nature is electron-like centered in $(0,0)$. Whereas, in the underdoped regime 
$(n_T=0.90$ in figure \ref{figure:FSh}(c)), the Fermi surface is hole-like centered in $(\pi,\pi)$.
Hence, there is a critical doping $x_c$ $(x_c\simeq 1-n_T^{(c)})$, where the Fermi 
surface changes its nature from electron- to hole-like. It has been verified experimentally
that some thermodynamic properties as the entropy and the magnetic susceptibility are greatly enhanced
with a peak in $x_c$ \cite{avella}. Moreover, also in $x_c$, the Hall coefficient $R_H$ reverses its sign \cite{avella}.
The figure \ref{figure:FSh}(d) shows the FS for $n_T=0.76$ and different values of hybridization.
Here, the hybridization has been considered ${\bf k}$-independent \cite{calegari2005EB} 
$(V_0^{pd})^2\equiv \langle V_{\bf k}^{dp}V_{\bf k}^{pd} \rangle$, where $\langle ...\rangle$ is the average over the Brillouin zone and $V_{\bf{k}}^{dp}(V_{\bf{k}}^{pd})$ are the Fourier transform of $t_{ij}^{dp}(t_{ij}^{pd})$.
It can be observed that the topology of the FS is directly associated with the hybridization $V_0^{pd}$. As a consequence, the value of $x_c$ depends on the hybridization. It is important to note that this dependence is pronounced in the points $(0,\pi)$ and $(\pi,0)$, where the intensity of the superconducting order parameter is maximum in the particular case of $d_{x^2-y^2}$-wave symmetry. In the present work, it has been found that this effect is related to a decreasing of the order parameter, as the hybridization increases.
Here, the quantity $V_0^{pd}$ is the {\bf k}-independent hybridization \cite{calegari2005EB}. 
The figure \ref{figure:FSh}(e) shows that the Coulomb interaction is also related to
a change in the FS topology.  
\begin{figure}[t!]
\begin{center}
\includegraphics[angle=-90,width=6.2cm]{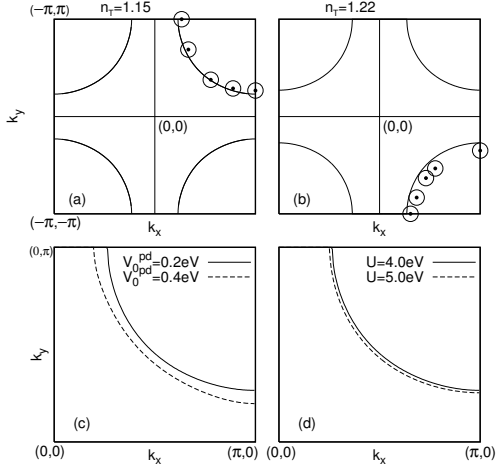}
\end{center}
\caption{(a) and (b) show the Fermi surface for different electron-doping. 
     The full lines show the result for $U=4$eV, $V_0^{pd}=0.2$eV and $k_BT=0.0011$eV.
     The remaining parameters are,
     $\varepsilon_{p}-\varepsilon_{d}=2.0$eV, $t^d=-0.3$eV, $t^p=-0.7$eV,
     $t^{ld}=0.02$eV and $t^{lp}=0$.  The 
     symbols $\odot$ correspond to the experimental data taken from Refs. 
     \cite{markiewicz2,King}.
     (c) Fermi surface for for $U=4$eV, $n_T=1.22$ and two different hybridizations. 
     (d) Fermi surface for $V_0^{pd}=0.2$eV, $n_T=1.22$ and two different values of Coulomb interaction.
}
\label{figure:SFe}
\end{figure}
The FS associated with the electron-doped case is shown in figure \ref{figure:SFe}. 
As it can be noted, the change in the topology of the FS upon doping does not occur.
It should be observed in figure \ref{figure:SFe}(c)
that, as in the hole-doped situation, the 
hybridization acts in the sense of changing the area enclosed by the 
FS. However, here, different of the hole-doped case,
the area enclosed by the FS decreases if the hybridization increases. 
It has been found that the effect of the Coulomb interaction on the Fermi surface is
similar to the hybridization effect. 

It is important to highlight that, in hole-doped systems, experimental results indicate that above $x_c$, where the 
Hall coefficient $R_H$ is negative, the superconductivity disappears \cite{ono}. The peculiar evolution of FS in the hole-doped regime and its relation with $R_H$ 
through $x_c$ suggest that, in cuprate systems, the effect of the hybridization is very important, at least in the hole-doped case.

\end{document}